\documentclass[aps,pre,floats,twocolumn,superscriptaddress]{revtex4}
\usepackage{latexsym,amsmath}
\usepackage{amsbsy}
\usepackage[usenames]{color}
\usepackage[pdftex]{graphicx}
\usepackage{amssymb}

\begin{document}

\title{Patterns of dominant flows in the world trade web}

\author{M. \'Angeles Serrano}
\affiliation{Institute of Theoretical Physics, LBS, FSB,
EPFL, BSP 725 - Unil, 1015 Lausanne, Switzerland}

\author{Mari{\'a}n Bogu{\~n}{\'a}}
\affiliation{Departament de F{\'\i}sica Fonamental, Universitat de
  Barcelona,\\ Mart\'{\i} i Franqu\`es 1, 08028 Barcelona, Spain}

\author{Alessandro Vespignani}
\affiliation{School of Informatics, Indiana University, Eigenmann
Hall, 1900 East Tenth Street, Bloomington, IN 47406, USA}
\affiliation{Complex Network Lagrange Laboratory (CNLL), Institute for Scientific Interchange (ISI), Torino 10133, Italy}

\date{today}

\begin{abstract}
The large-scale organization of the world economies is exhibiting
increasingly levels of local heterogeneity and global
interdependency. Understanding the relation between local and global
features calls for analytical tools able to uncover the global
emerging organization of the international trade network. Here we
analyze the world network of bilateral trade imbalances and
characterize its overall flux organization, unraveling local and
global high-flux pathways that define the backbone of the trade
system. We develop a general procedure capable to progressively
filter out in a consistent and quantitative way the dominant trade
channels. This procedure is completely general and can be applied to
any weighted network to detect the underlying structure of transport
flows. The trade fluxes properties of the world trade web determines
a ranking of trade partnerships that highlights global
interdependencies, providing information not accessible by simple
local analysis. The present work provides new quantitative tools for
a dynamical approach to the propagation of economic crises.
\end{abstract}

\maketitle

\section{Introduction}
The term ``globalization'', when applied to the international
economic order, refers to the presence of an intricate network of
economic partnership among an increasing number of
countries~\cite{Centeno:2006}. In this context, the International
trade system, describing the fundamental exchange of goods and
services, plays a central role as one of the most important
interaction channels between states~\cite{Krugman:1995}. For
instance, it broadly defines the substrate for the spreading of
major economic crises~\cite{Rose:1999,Forbes:2002,Forbes:2005}, such
as the 1997 Asiatic
crisis~\cite{Forbes:2005,Goldstein:1998,Wang:1998} which shows how
economic perturbations originated in a single country can somehow
propagate globally in the world. Moreover, commercial trade flows
are indeed highly correlated with other types of cross-country
economic interactions (flows of services, financial assets, workers)
and so stand as a good indicator for more general economic
relations~\cite{Krugman:2005}.
The International trade system as an independent extract of the world
economy is therefore still a partial view of the whole system;
a complete description would
consider the feedback mechanisms that
operate between international trade imbalances and other economic
variables such as investment, debt, or currency prices.  On the other
hand, the study of the International trade network in a system's
perspective represents a necessary first step
before proceeding with a subsequent more integrative investigation
and has proven to be successful in providing insight into some of
its global properties.

The large size and the entangled connectivity pattern characterizing
the international trade organization point out to a complex system
whose properties depend on its global structure.  In this
perspective, it appears natural to analyze the world trade system at
a global level, every country being important regardless of its size
or wealth and fully considering all the trade relationships. A
convenient framework for the analysis of complex interconnected
systems is network analysis~\cite{Barabasi:2002Rev,Mendesbook}.
Within this approach, countries are represented as nodes and trade
relationships among them as links. Such visualizations of bilateral
trade relations have been used in recent years to help analyze
gravity models~\cite{Plumper:1999,Plumper:2003}, often proposed to
account for the world trade patterns and their
evolution~\cite{Bergstrand:1985}. While the first attempts to study
the trade system as a complex network have successfully revealed a
hierarchical
organization~\cite{WTW,Garlaschelli:2004,Garlaschelli:2005}, these
studies focused on topological aspects neglecting fundamental
components, such as the heterogeneity in the magnitude of the
different bilateral trade relations and their asymmetry. These are
essential issues in the understanding of the interplay between the
underlying structure and the principles that rule the functional
organization of the system.

Here we tackle the quantitative study of the world trade network by
implementing the trade flux analysis at a global scale. To this end,
we construct the weighted directed network of merchandize trade
imbalances between world countries. In this representation, each
country appears as a node and a directed link is drawn among any
pair whenever a bilateral trade imbalance exists, i.e., whenever
bilateral imports does not balance exports. The direction of the
arrow follows that of the net flow of money and it is weighted
according to the magnitude of the imbalance between the two
countries. More precisely, we define the elements $E_{ij}$ that
measure the exports of country $i$ to country $j$ and the elements
$I_{ij}$ that measure the imports of country $i$ from country $j$.
The trade imbalance matrix is therefore defined as $T_{ij} =
E_{ij}-I_{ij}$ and measures the net money flow from country $j$ to
country $i$ due to trade exchanges. Since $E_{ij}=I_{ji}$ and
$I_{ij}=E_{ji}$, $T$ is an antisymmetric matrix with
$T_{ij}=-T_{ji}$, and a directed network can be easily constructed
by assuming a directed edge pointing to the country with positive
balance. The network of the net trade flows is therefore defined in
terms of a weighted adjacency matrix $F$ with $F_{ij}=\mid T_{ij}
\mid=\mid T_{ji}\mid$ for all $i,j$ with $T_{ij}<0$, and $F_{ij}=0$
for all $i,j$ with $T_{ij}\geq 0$ (see Fig.~1a for a pictorial
description).

\begin{figure}[t]
\begin{center}
\hspace{1.25cm}
\includegraphics[width=7cm]{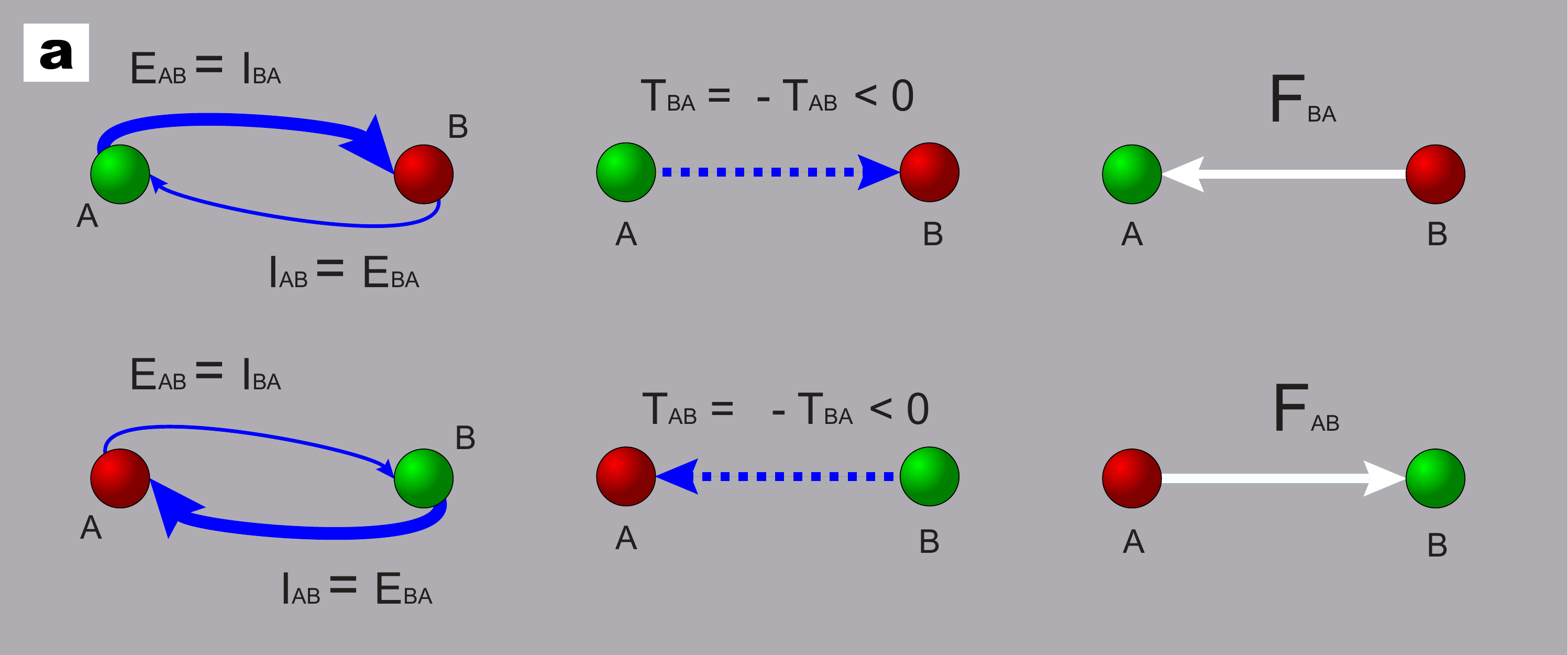}\\
\includegraphics[width=10cm]{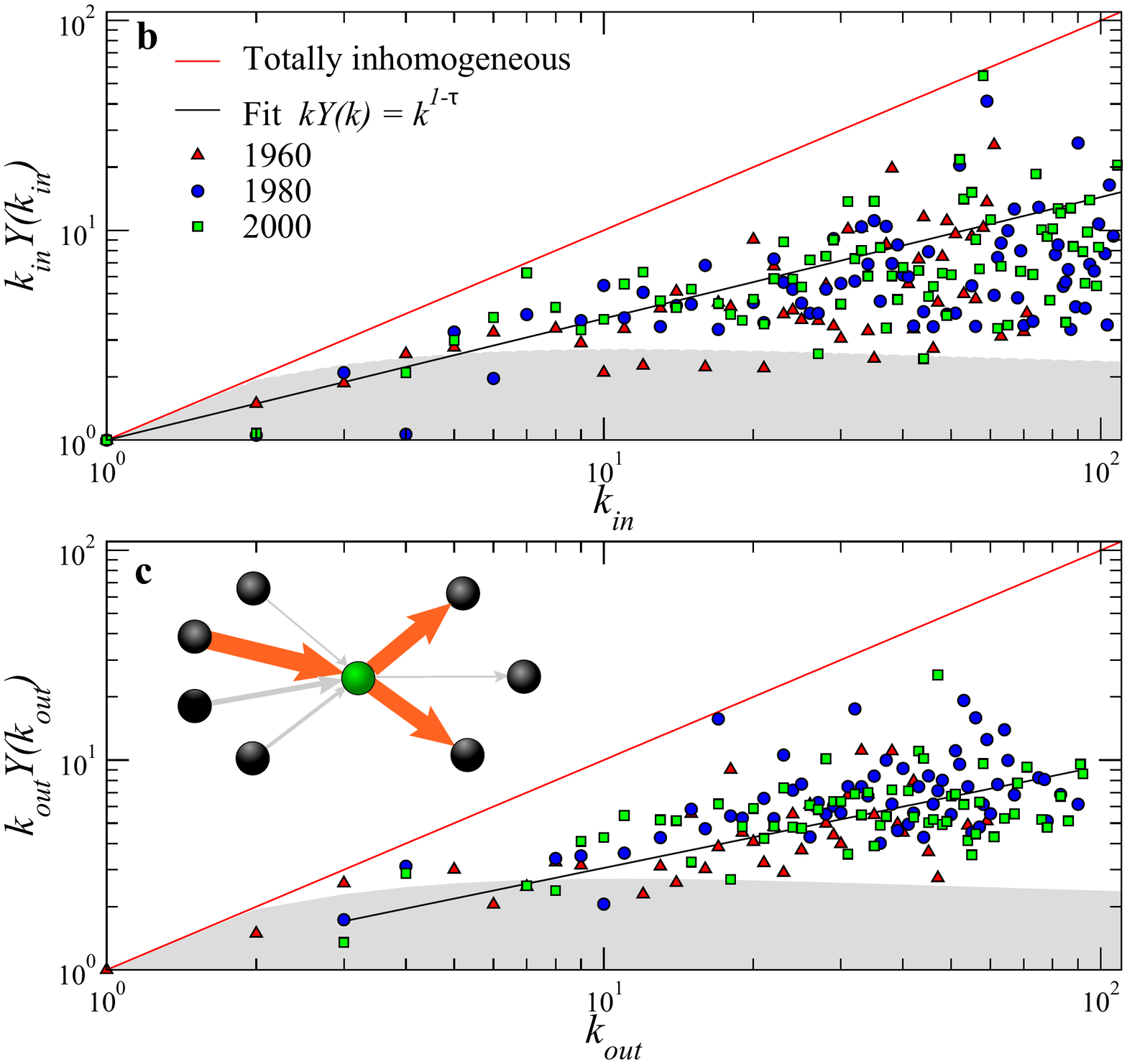}\\
\caption{Measuring local inhomogeneity in fluxes. {\bf a},
conceptual representation of the link construction process. {\bf b}
and {\bf c}, local inhomogeneity for incoming ({\bf b}) and outgoing
({\bf c}) connections measured by $kY(k)$ as compared to the null
model. The diagonal line corresponds to the maximum possible
inhomogeneity, with only one connection carrying all the flux. The
line $kY(k)=1$ is the maximum homogeneity, with all the fluxes
equally distributed among the connections. The area depicted in grey
corresponds to the average of $kY(k)$ under the null model plus two
standard deviations. The solid lines are the best fit estimates
which give $k_{in}Y(k_{in})\sim k_{in}^{0.6}$ and
$k_{out}Y(k_{out})\sim k_{out}^{0.5}$. The inset in ({\bf c})
sketches a pathway through a country arising from strong local
inhomogeneity in incoming and outgoing connections.}
 \label{fig:1}
\end{center}
\end{figure}

By using the above procedure we constructed the network of trade
imbalances by using the data set which reports the annual
merchandize trade activity between independent states in the world
during the period 1948-2000,
together with the annual values of their Gross Domestic Product per
capita and population figures (1950--2000)~\cite{Gleditsch:2002},\footnote{
(version (4.1),\\ http://weber.ucsd.edu/$\sim$kgledits/exptradegdp.html)
The following issues should be considered:
i) This expanded trade database includes additional estimates for missing
values.
ii) The definition of state in the international system is as
defined by the Correlates of War Project
(http://www.correlatesofwar.org/).
iii) The figures for trade flows are in millions of current-year US
dollars.
iv) The import/export values correspond to exchanges of merchandizes.}.
The time span of
the data set allows us to study the change of trade flow networks
with yearly snapshots characterizing the time evolution of the trade
system. The most basic topological
characterization of each country within the network is given by the
number of incoming and outgoing links, $k_{in}$ and $k_{out}$
respectively, which inform us about the number of neighboring
countries that contribute positively and negatively to the net trade
imbalance of the country under consideration. A precise assessment
of the country trade balance cannot however neglect the magnitude of
the fluxes carried by each trade relation. This information can be
retrieved summing up all the weights of the incoming or outgoing
links, which give us the total flux of money due to trade entering
to or leaving from the country of interest. In the network
literature, these two variables are called incoming and outgoing
strength and are denoted by $s_j^{in}=\sum_i F_{ij}$ and
$s_j^{out}=\sum_i F_{ji}$, respectively~\cite{Vespignani:2004WAN}.
The total trade imbalance of a country can then be computed as
$\Delta s_j=s_j^{in}-s_j^{out}$. Depending on $\Delta s_j$,
countries can be then defined as net consumers and net producers.
Net producers export more than they import, the total outcome being
a trade surplus which corresponds to $\Delta s_j>0$, whereas net
consumers export less than they import, the total outcome being a
trade deficit which is indicated by $\Delta s_j<0$. Since one
incoming link for a given country is always an outgoing link for
another, the sum of all the countries' trade imbalances in the
network must be zero. While the local balance is not conserved, we
are therefore dealing with a closed system which is globally
balanced (the total flux is conserved). Merchandizes, or
equivalently money, flows in the system from country to country with
the peculiarity that there is a global flow of money from consumer
countries to producer ones.

\section{Local heterogeneity and backbone extraction}
The obtained networks show a high density of connections and
heterogeneity of the respective fluxes among countries. Indeed, as
the number of countries increases, so does the average number of
trade partners, as well as the total flux of the system, which is
seen to grow proportional to the aggregated world Gross Domestic
Product~\cite{Serrano:2007}. The overall flux organization at the
global scale can be characterized by the study of the flux
distribution. A first indicator of the system heterogeneity is
provided by the probability distribution $P(F_{ij})$ denoting the
probability that any given link is carrying a flux $F_{ij}$. The
observed distribution is heavy-tailed and spans approximately four
orders of magnitude~\cite{Serrano:2007}. Such a feature implies that
only a small percentage of all the connections in the network carry
most of its total flow $F$ and that there is no characteristic flux
in the system, with most of the fluxes below the average and some of
them with a much higher value. This is however not totally
unexpected since a large scale heterogeneity is a typical feature of
large-scale networks. In addition, the global heterogeneity could
just be due to differences in the sizes of the countries, in their
population and in their respective Gross Domestic Product. More
interesting is therefore the characterization of the local
heterogeneity; i.e. given all the connections associated to each
given country, how is the flux distribution for each of them.

\begin{table}[t]
\begin{center} \caption{Sizes of the backbones. Percentage of the original total weight $F$, number
of nodes $N$ and links $E$ in the 1960 and 2000 imbalance networks
that remain in the backbone as a function of the significance level
$\alpha$.}
\begin{tabular}{lcccccc}
\hline
&\multicolumn{3}{c}{\hspace{0.7cm} 1960}&\multicolumn{3}{c}{\hspace{0.7cm} 2000}\\
\hline \hspace{0.5cm}$\alpha$ &\hspace{0.7cm}\%F&\hspace{0.05cm}\%N&\hspace{0.05cm}\%E\mbox{\hspace{0.05cm}}&\hspace{0.7cm}\%F&\hspace{0.05cm}\%N&\hspace{0.05cm}\%E \mbox{\hspace{0.05cm}}\\
\hline \hspace{0.5cm}0.2&\hspace{0.7cm}88&\hspace{0.05cm}100&\hspace{0.05cm}25\mbox{\hspace{0.05cm}}&\hspace{0.7cm}92&\hspace{0.05cm}98&\hspace{0.05cm}25\mbox{\hspace{0.05cm}}\\
 \hspace{0.5cm}0.1&\hspace{0.7cm}83&\hspace{0.05cm}100&\hspace{0.05cm}19\mbox{\hspace{0.05cm}}&\hspace{0.7cm}87&\hspace{0.05cm}98&\hspace{0.05cm}19\mbox{\hspace{0.05cm}}\\
 \hspace{0.5cm}0.05&\hspace{0.7cm}79&\hspace{0.05cm}99&\hspace{0.05cm}15\mbox{\hspace{0.05cm}}&\hspace{0.7cm}84&\hspace{0.05cm}97&\hspace{0.05cm}15\mbox{\hspace{0.05cm}}\\
 \hspace{0.5cm}0.01&\hspace{0.7cm}69&\hspace{0.05cm}92&\hspace{0.05cm}9\mbox{\hspace{0.05cm}}&\hspace{0.7cm}75&\hspace{0.05cm}96&\hspace{0.05cm}10\mbox{\hspace{0.05cm}}\\
\hline
\end{tabular}
\end{center}
\label{table_conf}
\vspace{-0.5cm}
\end{table}

A local heterogeneity implies that only a few links carry the
biggest proportion of the country's total in-flow or out-flow.
Interestingly, such a heterogeneity would define specific pathways
within the network that accumulate most of the total flux. In order
to asses the effect of inhomogeneities at the local level, for each
country $i$ with $k$ incoming or outgoing trade partners we
calculate~\cite{Guichard:2003,Almaas:2004}
\begin{equation}
kY_i(k)= k\sum_{j=1}^{k}p_{ij}^2, \label{eq:Y}
\end{equation}
where $k$ can be either $k_{in}$ or $k_{out}$ in order to discern
between inhomogeneities in incoming and outgoing fluxes, and where
the normalized fluxes of node $i$ with its neighbors are calculated
as $p_{ij}=F_{ji}/s^{in}_i$ for incoming connections and as
$p_{ij}=F_{ij}/s^{out}_i$ for the outgoing ones. The function
$Y_i(k)$ is extensively used in economics as a standard indicator of
market concentration, referred as the Herfindahl-Hirschman Index or
HHI~\cite{HHIHerfindahl:1959,HHIHirschman:1964}, and it was also
introduced in the complex networks literature as the disparity
measure~\cite{Derrida:1987}. In all cases, $Y_i(k)$ characterizes
the level of local heterogeneity. If all fluxes emanating from or
arriving to a certain country are of the same magnitude, $kY_i(k)$
scales as 1 independently of $k$, whereas this quantity depends
linearly  on $k$ if the local flux is heterogeneously organized with
a few main directions. Increasing deviations from the constant
behavior are therefore indicating heterogeneous situations in which
fluxes leaving or entering each country are progressively peaked on
a small number of links with the remaining connections carrying just
a small fraction of the total trade flow. On the other hand, the
deviations from the constant behavior have to be expected for low
values of $k$ and it is important to compare the obtained results
with the deviations simply produced by statistical fluctuations. To
this end, we introduce a null model for the distribution of flows
among a given number of neighbors in order to assess, in a case by
case basis, whether the observed inhomogeneity can just be due to
fluctuations or it is really significant.

The null model with the maximum random homogeneity corresponds to
the process of throwing $k-1$ points in a $[0,1]$ interval, so that
the interval ends up divided in k sections of different lengths
representing the different values assigned to the $k$ variables
$p_{ij}$ in the random case. It can be analytically proved that the
probability that one of these variables takes a particular value $x$
depends on $k$ and is
\begin{equation}
Prob\{x<p_{ij}<x+dx\}=(k-1)(1-x)^{k-2}dx. \label{eq:PX}
\end{equation}
This probability density function can be used to calculate the
statistics of $kY_{NM}(k)$ for the null model. Both the average and
the standard deviation are found to depend on $k$:
\begin{equation}
\langle k Y_{NM}(k)\rangle = k \langle Y_{NM}(k)\rangle = \frac{2k}{k+1}
\end{equation}
\begin{equation}
\sigma^2\left(kY_{NM}(k)\right)=  k^2
\left(\frac{20+4k}{(k+1)(k+2)(k+3)}-\frac{4}{(k+1)^2}\right),
\label{eq:NMY}
\end{equation}
so that each node in the network with a certain in or out degree
should be compared to the corresponding null model depending on the
appropriate $k$.

In Fig.~1, we show the empirical measures along with the region
defined by the average value of the same quantity $kY(k)$ plus two
standard deviations as given by the null model (shadowed area in
grey). For a homogeneously random assignment of weights, this
quantity converges to a constant value for large $k$, which is
clearly different from the observed empirical behavior. Most
empirical values lie out of the null model domain, which proves that
the observed heterogeneity is due to a well definite ordering
principle and not to random fluctuations.

The direct fit of the data indicates that both in and out fluxes
follow the scaling law $kY_i(k)\propto k^{\beta}$ with
$\beta_{in}=0.6$ for the incoming connections and $\beta_{out}=0.5$
for the outgoing ones (see Fig.~1). This scaling represents and
intermediate behavior between the two extreme cases of perfect
homogeneity or heterogeneity but clearly points out the existence of
strong local inhomogeneities. The emerging picture is therefore consistent
with the existence of major pathways of trade flux imbalances (thus
money) that enters the country using its major incoming links and
leaves it through its most inhomogeneous outgoing trade channels
(see inset in Fig.~1{\bf c}).

\begin{figure*}[t]
\begin{center}
 \includegraphics[width=12cm]{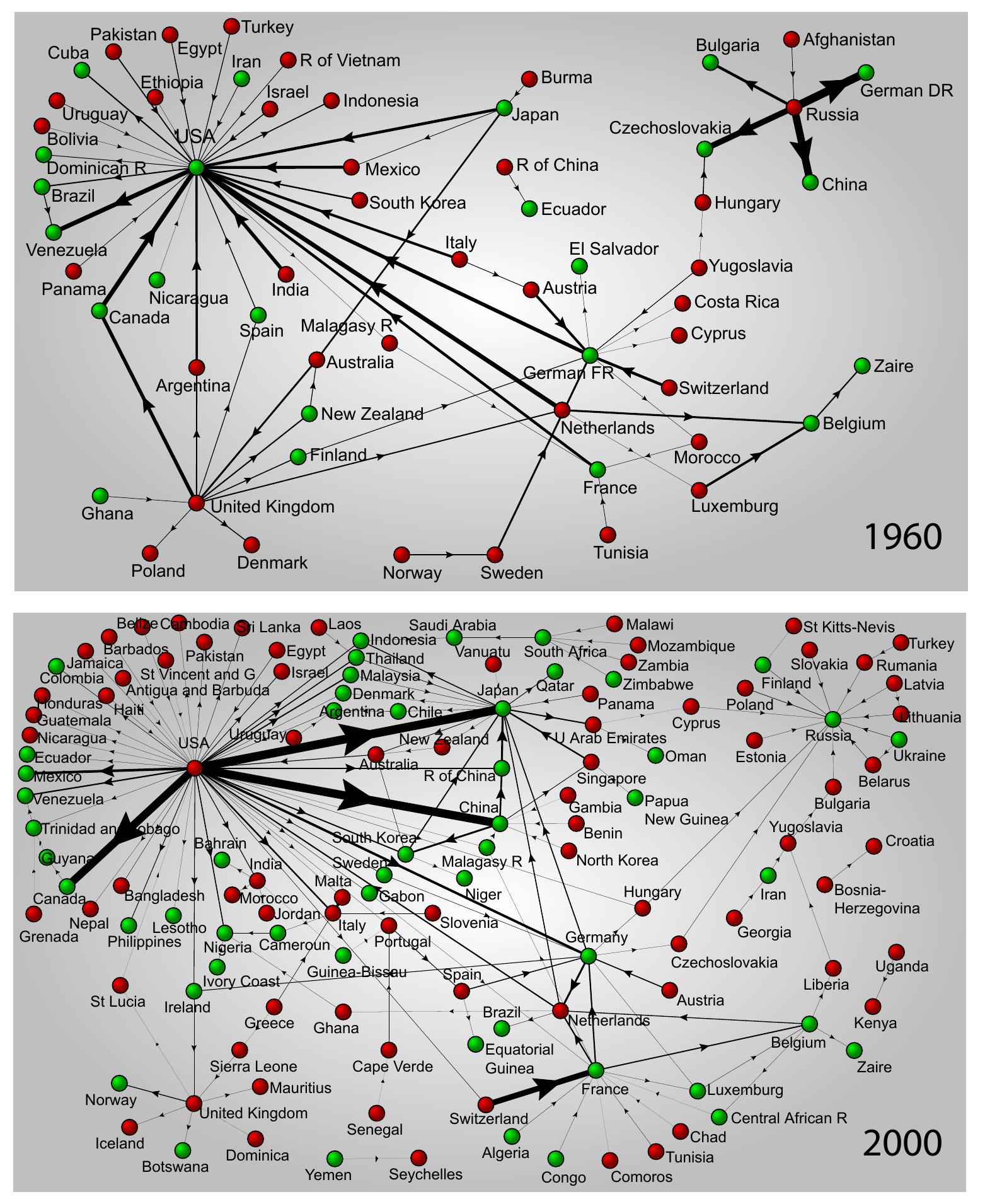}
\end{center}
\vspace{-0.5cm}
  \caption{Backbone of the world trade system. Snapshots of the $\alpha=10^{-3}$ backbone of the world trade
  imbalance web for the years 1960 and 2000. Notice that the most central economies
  are depicted at fixed positions to make both graphs more easily comparable.}
  \label{fig:evolution}
\end{figure*}

The analysis of the local inhomogeneities in the trade fluxes
prompts to the presence of high-flux backbones, sparse subnetworks
of connected trade fluxes carrying most of the total flux in the
network. This backbone is necessarily encoding a wealth of
information being the dominating structure of the trade system. It
is also worth remarking that the local heterogeneity is not just
encoded in high flux links in terms of their absolute intensities,
but also takes into account the local heterogeneity by comparing the
strength of the fluxes associated to a given country with its total
strength. It is then interesting to filter out this special links
and provide snapshots of the trade system backbone. This can be
achieved by comparing the link fluxes with the null model used for
the calculation of the disparity in a pure random case. The same
approach allows us the calculation for each connection
of a country $i$ of the probability $\alpha_{ij}$ that its
normalized flux value $p_{ij}$ is due to chance. Along these lines, we
can identify highly
inhomogeneous fluxes as those which satisfy
\begin{equation}
\alpha_{ij}=1-(k-1)\int_0^{p_{ij}} (1-x)^{k-2}dx < \alpha,
\label{eq:confidencelevel}
\end{equation}
where $\alpha$ is a fixed significance level. Notice that this expression
depends on the number of connections of each country,
$k$. By choosing a global threshold for all countries we obtain a
homogeneous criteria that allows us to compare inhomogeneities in
countries with different number of connections and filter out links that carry
fluxes which can be considered not compatible with a random
distribution with an increasing statistical confidence. The backbone is then
obtained by preserving all the links which beat the threshold for at
least one of the two countries at the ends of the link while
discounting the rest. By changing the
significance level we can filter out the links focusing on
progressively more relevant heterogeneities and backbones.

An important aspect of this new filtering algorithm is that it does
not belittle small countries and then, it offers a systematic
procedure to reduce the number of connections without diminishing
the number of countries and choosing the backbone according to the
amount of trade flow we intend to characterize. It provides a
quantitative and consistent way to progressively identify the
relevant flow backbone once the level of statistical confidence with
respect to the null case is fixed, or instead the total flow we want
to represent in the system. Indeed, it is remarkable that when
looking at the network of the year 2000 one finds that the
$\alpha=0.05$ backbone contains only $15\%$ of the original links
yet accounting for $84\%$ of the total trade imbalance. Most of the
backbones form a giant connected component containing most of the
countries in the network, and only for very high values of the
confidence level, defining a sort of {\em super-backbones}, some
disconnected components appear and the number of countries starts to
drop.  In this respect, the $\alpha=0.01$ backbone seems to offer
the best trade-off since it keeps nearly all countries, $75\%$ of
the total trade imbalances, and one order of magnitude less
connections than in the original network (see Table~1).

The backbone reduction is extremely effective in sorting out the
most relevant part of the network and can be conveniently used for
visualization purposes. For the sake of space and reproduction
clarity, we report the backbones corresponding to $\alpha=10^{-3}$,
still accounting for approximately $50\%$ of the total flux of the
system. Fig.~2 shows two snapshots of such backbones for 1960 and
2000. These high-flux backbones evidence geographical, political and
historical relationships among countries which affect the observed
trade patterns. For instance, the trade of US with its
geographically closer neighbors and also continental neighbors, the
case of Russia and the former Soviet republics, or France and its
former colonies, the lack of strong trade relations between the two
blocks in the cold war, more evident in 1960. In general terms, a
recurrent motif present in all years is the star-like structure,
formed by a central powerful economy surrounded by small dependent
economies. The USA appears as one of those powerful hubs during all
this period. However, other countries has gradually lost this role
in favor of others. This is the case of the UK, which was the only
star-like counterpart of the USA in 1948; since then its direct
area of influence has been narrowing. On the contrary, other
countries have arisen for different reasons as new hub economies.
This is the case of some European countries, Japan, and most
recently, China.

\begin{table*}[t]
\caption{Rankings from the Dollar experiment. Top: effect of two
major source countries, USA and Switzerland, on the rest of the
world. The first list is a top ten ranking of countries according to
$e_{ij}$, where the index $i$ stands for the analyzed source. The
second list is the top ten ranking of direct bilateral trade
measured as the percentage of flux from the source country, that is,
$e_{ij}^{local}=F_{ij}/s^{out}_i$. Bottom: major contributors to two
major sink countries, Japan and Russia. The first list is a top ten
ranking of countries according to $g_{ij}$, $i$ standing for the
analyzed sink. The second list is the top ten ranking due to direct
trade. In this case, the direct contribution is
$g_{ij}^{local}=F_{ji}/s^{in}_i$. Countries in boldface have no
direct connection with the analyzed country. The values for $e_{ij}$
and $g_{ij}$ are obtained from the simulation of the dollar
experiment described in the text using $10^6$ different realizations
for each country, for the year 2000.}
\begin{center}
\begin{tabular}{|c|c|c|c||c|c|c|c|} \hline
\multicolumn{8}{|c|}{Net Consumers - Sources} \\ \hline
\multicolumn{4}{|c||}{USA}&\multicolumn{4}{|c|}{Switzerland}\\
\hline \multicolumn{2}{|c|} {Dollar experiment}
&\multicolumn{2}{|c||} {Bilateral trade}&
\multicolumn{2}{|c|}{Dollar experiment} &
\multicolumn{2}{|c|}{Bilateral trade}\\
\hline
Japan            & 19.5\%  &Japan           &17.2\%     &France                 &27.3\%    &France&75.0\% \\
Canada           & 9.9\%   &China           &16.7\%     &Germany                &10.0\%    &Germany   &9.5\%\\
China            & 9.3\%   &Canada          &15.6\%     &Russia                 &9.7\%     &Russia    &4.1\% \\
Saudi Arabia     & 6.1\%   &Mexico          &5.1\%      &\textbf{Japan}            &8.5\%     &Netherlands   &2.6\% \\
Russia           & 5.4\%   &Germany         &4.8\%      &Ireland                &6.9\%     &Ireland     &2.3\% \\
Germany          & 4.5\%   &R of China      &3.1\%      &\textbf{Norway}           &6.0\%     &Belgium   &1.7\% \\
Indonesia        & 4.3\%   &Italy           &3.0\%      &\textbf{Saudi Arabia}     &4.2\%     &Italy   &1.2\% \\
Malaysia         & 3.9\%   &Venezuela       &2.8\%      &\textbf{China}            &3.4\%     &Austria   &1.1\% \\
Ireland          & 2.7\%   &South Korea     &2.4\%      &\textbf{Indonesia}        &2.3\%     &Libya    &0.4\% \\
South Korea      & 2.7\%   &Malaysia        &2.4\%      &\textbf{Malaysia}        &1.9\%     &Nigeria   &0.4\% \\
\hline\hline \multicolumn{8}{|c|}{Net Producers - Sinks} \\
\hline
\multicolumn{4}{|c||}{Japan}&\multicolumn{4}{|c|}{Russia}\\
\hline \multicolumn{2}{|c|}{Dollar experiment} &
\multicolumn{2}{|c||}{Bilateral trade} &\multicolumn{2}{|c|}{Dollar
experiment}& \multicolumn{2}{|c|} {Bilateral trade}\\ \hline
USA                    &62.6\%  &USA             &40.2\%   &USA                    &33.3\%  &Germany            &9.0\%  \\
UK                     &7.3\%   &R of China      &9.3\%    &UK                     &7.2\%   &Italy              &8.1\%  \\
Spain                  &3.8\%   &Singapore       &7.0\%    &Switzerland            &7.1\%   &USA                &7.7\%  \\
\textbf{Switzerland}    &3.3\%   &South Korea     &5.6\%    &Poland                 &7.0\%   &China              &5.9\%  \\
Singapore              &2.4\%   &Germany         &5.1\%    &Turkey                 &6.9\%   &Poland             &5.4\%  \\
Turkey                 &2.1\%   &UK              &4.8\%    &Spain                  &5.1\%   &Japan              &4.4\%  \\
Panama                 &2.1\%   &Netherlands     &4.8\%    &Greece                 &3.5\%   &Turkey             &4.3\%  \\
Greece                 &1.9\%   &China           &3.9\%    &Egypt                  &2.2\%   &Switzerland        &4.0\%  \\
Portugal               &1.5\%   &Mexico          &2.1\%    &Lithuania              &2.0\%   &Netherlands        &4.0\%  \\
Egypt                  &1.5\%   &Thailand        &2.1\%    &Portugal               &1.9\%   &UK                 &3.6\%  \\
\hline
\end{tabular}
\end{center}
\label{table_sources}

\end{table*}

\section{Diffusion on complex networks and the Dollar experiment}
The picture emerging from our analysis has intriguing similarities
with other directed flow networks, such as metabolic
networks~\cite{Almaas:2004}, that transport information, energy or
matter. Indeed, the trade imbalances network can be seen as a
directed flow network that transport money. In other words, we can
imagine that net consumer countries are injecting money in the
system. Money flows along the edges of the network to finally reach
producer countries. Producer countries, however, do not absorb
completely the incoming flux, redistributing part of it through the
outgoing links. The network is therefore characterizing a complex
dynamical process in which the net balance of incoming and outgoing
money is the outcome of a global diffusion process. The realization
of such a non-local dynamics in the flow of money due to the trade
imbalances spurs the issue of what impact this feature might have on
the effect that one economy can have on another. In order to tackle
this issue we perform a simple numerical study, defined as the
``{\em dollar experiment}''. The ``experiment'' considers running on
the networks two symmetric random walk processes. Since we are
limited by the yearly frequency of the empirical data, we assume at
first approximation that the time scale of the changes in the
structure of the underlying trade imbalances network is bigger than
the characteristic diffusion time of the random walk processes. In
the first case we imagine that a consumer country ($\Delta s <0$) is
injecting one dollar from its net debit into the system. The dollar
at this point travels through the network following fluxes chosen
with a probability proportional to their intensity, and has as well
a certain probability of being trapped in producer countries
($\Delta s >0$) with a probability $P_{abs}=\frac{\Delta
s}{s_{in}}$. More precisely, if we consider a consumer country, such
as the USA, the traveling dollar goes from country to country always
following outgoing fluxes chosen with a probability proportional to
their intensity. If in its way it finds another source it just
crosses it, whereas if it finds a producer country $j$ it has a
probability $P_{abs}(j)$ of being absorbed. Mathematically, this
process is a random walk on a directed network with heterogeneous
diffusion probability and in the presence of sinks. By repeating
this process many times it is possible to obtain the probability
$e_{ij}$ that the traveling dollar originated in the source $i$ is
finally absorbed in the sink $j$. In other words, for each dollar
that a source country $i$ adds to the system, $e_{ij}$ represents
the fraction of that dollar that is retained in country $j$. The
symmetric process considers that each producer country is receiving
a dollar and the traveler dollar goes from country to country always
following incoming links backward chosen with a probability
proportional to their intensity. If in its way it finds another sink
it just crosses it, whereas if it finds a source $j$ it has a
probability $P_{abs}(j)= \frac{\mid\Delta s \mid}{s_{out}}$ of
remaining in that country. The iteration of this process gives the
probability $g_{ij}$ that yields the fraction originated in the
source country $j$ of each dollar that a sink country retains.
Consequently, these two
quantities are related by the detailed balance condition
\begin{equation}
| \Delta s_{i}| e_{ij}=\Delta s_{j} g_{ji}.
\end{equation}
The matrices $e_{ij}$ and $g_{ji}$ are normalized probability
distributions and, therefore, they satisfy that
$\sum_{j;sink}e_{ij}=1$ and $\sum_{i;source}g_{ji}=1$. Using this
property in the detailed balance condition, we can write
\begin{equation}
\Delta s_{j}=\sum_{i:source} e_{ij} |\Delta s_{i}| \mbox{
\hspace{0.4cm} and \hspace{0.4cm} } |\Delta s_{i}|=\sum_{j:sink}
g_{ji} \Delta s_{j}.
\end{equation}
Then, the total trade imbalance of a sink or source country can be
written as a linear combination of the trade imbalances of the rest
of the source or sink countries, respectively. Therefore, by measuring $e_{ij}$, it is
possible to discriminate the effect that one economy has on another
or, with $g_{ij}$, to find out which consumer country is
contributing the most to a producer one, in both cases taking into
account the whole topology of the network and the inhomogeneities of
the fluxes. The advantage of this approach lies on its simplicity
and the lack of tunable parameters. Indeed, all the information is
contained in the network itself, without assuming any kind of
modeling on the influences among countries.

By using this experiment it is possible to evaluate for a consumer
country where the money spent is finally going. For each dollar
spent we know which percentage is going to any other producer
country and we can rank those accordingly. It is important to remark
that in this case countries  might not be directly connected since
the money flows along all possible paths, sometimes through
intermediate countries. This kind of ranking is therefore different
from the customarily considered list of the first neighbors ranked
by magnitude of fluxes. The analysis indeed shows unexpected results
and, as it has been already pointed out in other
works~\cite{Forbes:2005} applying other methodologies, a country can
have a large impact on other countries despite being a  minor or
undirect trading partner, see Table~2. Similarly, producer countries
may have a share of the expenditure of non directly connected
countries resulting in a very different ranking of their creditors.
As an example, for each net dollar that the USA inject into the
system, only $9.3\%$ is retained in China although the direct
connection imbalance between these countries is $16.7\%$. Very
interestingly, we find that Switzerland spend a large share of his
trade imbalance in countries which do not have appreciable trade
with it and are therefore not directly connected such as Japan,
Indonesia, and Malaysia. The Swiss dollars go to these countries
after a long path of trade exchanges mediated by other countries. By
focusing on producer countries we find other striking evidence.
While the first importer from Russia by looking locally at the
ranking of trade imbalances is Germany, the global analysis shows
that one third of all the money Russia gains from trade is coming
directly or undirectly from the USA. In Table~2, we report other
interesting anomalies detected by the global analysis.

\section{Conclusions}
In summary, we have introduced a novel quantitative approach
applicable to any dense weighted complex network which filters out
the dominant backbones while preserving most of the nodes in the
original connected component. We have also discussed the behavior of
a coupled dynamical process, the dollar experiment, which unveils
the global properties of economic and trade partnerships. In a
globalized economy, we face ever increasing problems in
disentangling the complex set of relations and causality that might
lead to crisis or increased stability. Focusing on just the
bilateral relations among country economies is a reductionist
approach that cannot work in a highly interconnected complex
systems. We have proposed the use of the trade network
representation and mathematical tools that allow to uncover some
basic ordering emerging from the global behavior and the inclusion
of non-local effects in the analysis of trade interdependencies.
Future work on this grounds might help in the assessment of world
trade relations and the understanding of the global dynamics
underlying major economic crises.

\begin{acknowledgments}
We thank F. Vega-Redondo for useful comments.
M. B. acknowledges financial support by DGES grant No.
FIS2004-05923-CO2-02 and Generalitat de Catalunya grant No.
SGR00889. A.V. is partially supported by the NSF award IIS-0513650.
\end{acknowledgments}


\begin{thebibliography}{24}
\expandafter\ifx\csname natexlab\endcsname\relax\def\natexlab#1{#1}\fi
\expandafter\ifx\csname bibnamefont\endcsname\relax
  \def\bibnamefont#1{#1}\fi
\expandafter\ifx\csname bibfnamefont\endcsname\relax
  \def\bibfnamefont#1{#1}\fi
\expandafter\ifx\csname citenamefont\endcsname\relax
  \def\citenamefont#1{#1}\fi
\expandafter\ifx\csname url\endcsname\relax
  \def\url#1{\texttt{#1}}\fi
\expandafter\ifx\csname urlprefix\endcsname\relax\def\urlprefix{URL }\fi
\providecommand{\bibinfo}[2]{#2}
\providecommand{\eprint}[2][]{\url{#2}}

\bibitem[{\citenamefont{Centeno et~al.}(2006)\citenamefont{Centeno, Cooke, and
  Curran}}]{Centeno:2006}
\bibinfo{author}{\bibfnamefont{M.~A.} \bibnamefont{Centeno}},
  \bibinfo{author}{\bibfnamefont{A.}~\bibnamefont{Cooke}}, \bibnamefont{and}
  \bibinfo{author}{\bibfnamefont{S.~R.} \bibnamefont{Curran}},
  \emph{\bibinfo{title}{NetMap Combined Studies, Mapping Globalization
  Project}} (\bibinfo{publisher}{Princeton University and University of
  Washington}, \bibinfo{year}{2006}).

\bibitem[{\citenamefont{Krugman}(1995)}]{Krugman:1995}
\bibinfo{author}{\bibfnamefont{P.~R.} \bibnamefont{Krugman}},
  \bibinfo{journal}{Brookings Papers on Economic Activity}
  \textbf{\bibinfo{volume}{1995}}, \bibinfo{pages}{327} (\bibinfo{year}{1995}).

\bibitem[{\citenamefont{Glick and Rose}(1999)}]{Rose:1999}
\bibinfo{author}{\bibfnamefont{R.}~\bibnamefont{Glick}} \bibnamefont{and}
  \bibinfo{author}{\bibfnamefont{A.}~\bibnamefont{Rose}}, \bibinfo{journal}{J.
  of Intl. Money and Finance} \textbf{\bibinfo{volume}{18}},
  \bibinfo{pages}{603} (\bibinfo{year}{1999}).

\bibitem[{\citenamefont{Forbes}(2002)}]{Forbes:2002}
\bibinfo{author}{\bibfnamefont{K.}~\bibnamefont{Forbes}}, in
  \emph{\bibinfo{booktitle}{In Sebastian Edwards and Jeffrey Frankel (eds.),
  Preventing Currency Crises in Emerging Markets}}
  (\bibinfo{organization}{University of Chicago Press, Chicago},
  \bibinfo{year}{2002}), pp. \bibinfo{pages}{77--124}.

\bibitem[{\citenamefont{Abeysinghe and Forbes}(2005)}]{Forbes:2005}
\bibinfo{author}{\bibfnamefont{T.}~\bibnamefont{Abeysinghe}} \bibnamefont{and}
  \bibinfo{author}{\bibfnamefont{K.}~\bibnamefont{Forbes}},
  \bibinfo{journal}{Rev. of Intl. Econ.} \textbf{\bibinfo{volume}{13}},
  \bibinfo{pages}{356} (\bibinfo{year}{2005}).

\bibitem[{\citenamefont{Goldstein}(1998)}]{Goldstein:1998}
\bibinfo{author}{\bibfnamefont{M.}~\bibnamefont{Goldstein}},
  \emph{\bibinfo{title}{The Asian Financial Crisis: Causes, Cures and Systemic
  Implications}} (\bibinfo{publisher}{International Institute for Economics},
  \bibinfo{address}{Washington, D.~C.}, \bibinfo{year}{1998}).

\bibitem[{\citenamefont{Noland et~al.}(1998)\citenamefont{Noland, Liu,
  Robinson, and Wang}}]{Wang:1998}
\bibinfo{author}{\bibfnamefont{M.}~\bibnamefont{Noland}},
  \bibinfo{author}{\bibfnamefont{L.-G.} \bibnamefont{Liu}},
  \bibinfo{author}{\bibfnamefont{S.}~\bibnamefont{Robinson}}, \bibnamefont{and}
  \bibinfo{author}{\bibfnamefont{Z.}~\bibnamefont{Wang}},
  \emph{\bibinfo{title}{Global Economics Effects of the Asian Currency
  Devaluations}} (\bibinfo{publisher}{International Institute for Economics},
  \bibinfo{address}{Washington, D.~C.}, \bibinfo{year}{1998}).

\bibitem[{\citenamefont{Krugman and Obstfeld}(2005)}]{Krugman:2005}
\bibinfo{author}{\bibfnamefont{P.~R.} \bibnamefont{Krugman}} \bibnamefont{and}
  \bibinfo{author}{\bibfnamefont{M.}~\bibnamefont{Obstfeld}},
  \emph{\bibinfo{title}{International Economics: Theory and Policy, Seventh
  Edition}} (\bibinfo{publisher}{Addison-Wesley}, \bibinfo{address}{Lebanon,
  Indiana, U.S.A.}, \bibinfo{year}{2005}).

\bibitem[{\citenamefont{Albert and Barab{\'a}si}(2002)}]{Barabasi:2002Rev}
\bibinfo{author}{\bibfnamefont{R.}~\bibnamefont{Albert}} \bibnamefont{and}
  \bibinfo{author}{\bibfnamefont{A.-L.} \bibnamefont{Barab{\'a}si}},
  \bibinfo{journal}{Reviews of Modern Physics 74, 47}
  \textbf{\bibinfo{volume}{74}}, \bibinfo{pages}{47} (\bibinfo{year}{2002}).

\bibitem[{\citenamefont{Dorogovtsev and Mendes}(2003)}]{Mendesbook}
\bibinfo{author}{\bibfnamefont{S.~N.} \bibnamefont{Dorogovtsev}}
  \bibnamefont{and} \bibinfo{author}{\bibfnamefont{J.~F.~F.}
  \bibnamefont{Mendes}}, \emph{\bibinfo{title}{Evolution of networks: From
  biological nets to the Internet and WWW}} (\bibinfo{publisher}{Oxford
  University Press}, \bibinfo{address}{Oxford}, \bibinfo{year}{2003}).

\bibitem[{\citenamefont{Krempel and Pl{\"{u}}mper}(1999)}]{Plumper:1999}
\bibinfo{author}{\bibfnamefont{L.}~\bibnamefont{Krempel}} \bibnamefont{and}
  \bibinfo{author}{\bibfnamefont{T.}~\bibnamefont{Pl{\"{u}}mper}},
  \bibinfo{journal}{J. of World-Systems Research} \textbf{\bibinfo{volume}{5}},
  \bibinfo{pages}{487} (\bibinfo{year}{1999}).

\bibitem[{\citenamefont{Krempel and Pl{\"{u}}mper}(2003)}]{Plumper:2003}
\bibinfo{author}{\bibfnamefont{L.}~\bibnamefont{Krempel}} \bibnamefont{and}
  \bibinfo{author}{\bibfnamefont{T.}~\bibnamefont{Pl{\"{u}}mper}},
  \bibinfo{journal}{J. of Social Structures} \textbf{\bibinfo{volume}{4}},
  \bibinfo{pages}{1} (\bibinfo{year}{2003}).

\bibitem[{\citenamefont{Bergstrand}(1985)}]{Bergstrand:1985}
\bibinfo{author}{\bibfnamefont{J.~H.} \bibnamefont{Bergstrand}},
  \bibinfo{journal}{The Review of Economics and Statistics}
  \textbf{\bibinfo{volume}{67}}, \bibinfo{pages}{474} (\bibinfo{year}{1985}).

\bibitem[{\citenamefont{Serrano and Bogu{\~n}\'{a}}(2003)}]{WTW}
\bibinfo{author}{\bibfnamefont{M.~A.} \bibnamefont{Serrano}} \bibnamefont{and}
  \bibinfo{author}{\bibfnamefont{M.}~\bibnamefont{Bogu{\~n}\'{a}}},
  \bibinfo{journal}{Phys. Rev. E} \textbf{\bibinfo{volume}{68}},
  \bibinfo{pages}{015101(R)} (\bibinfo{year}{2003}).

\bibitem[{\citenamefont{Garlaschelli and Loffredo}(2004)}]{Garlaschelli:2004}
\bibinfo{author}{\bibfnamefont{D.}~\bibnamefont{Garlaschelli}}
  \bibnamefont{and} \bibinfo{author}{\bibfnamefont{M.~I.}
  \bibnamefont{Loffredo}}, \bibinfo{journal}{Phys. Rev. Lett.}
  \textbf{\bibinfo{volume}{93}}, \bibinfo{pages}{188701}
  (\bibinfo{year}{2004}).

\bibitem[{\citenamefont{Garlaschelli and Loffredo}(2005)}]{Garlaschelli:2005}
\bibinfo{author}{\bibfnamefont{D.}~\bibnamefont{Garlaschelli}}
  \bibnamefont{and} \bibinfo{author}{\bibfnamefont{M.~I.}
  \bibnamefont{Loffredo}}, \bibinfo{journal}{Physica A}
  \textbf{\bibinfo{volume}{355}}, \bibinfo{pages}{138–144}
  (\bibinfo{year}{2005}).

\bibitem[{\citenamefont{Gleditsch}(2002)}]{Gleditsch:2002}
\bibinfo{author}{\bibfnamefont{K.~S.} \bibnamefont{Gleditsch}},
  \bibinfo{journal}{J. Conflict Resolut.} \textbf{\bibinfo{volume}{46}},
  \bibinfo{pages}{712–724} (\bibinfo{year}{2002}).

\bibitem[{\citenamefont{Barrat et~al.}(2004)\citenamefont{Barrat,
  Barth\'{e}lemy, Pastor-Satorras, and Vespignani}}]{Vespignani:2004WAN}
\bibinfo{author}{\bibfnamefont{A.}~\bibnamefont{Barrat}},
  \bibinfo{author}{\bibfnamefont{M.}~\bibnamefont{Barth\'{e}lemy}},
  \bibinfo{author}{\bibfnamefont{R.}~\bibnamefont{Pastor-Satorras}},
  \bibnamefont{and}
  \bibinfo{author}{\bibfnamefont{A.}~\bibnamefont{Vespignani}},
  \bibinfo{journal}{Proc. Natl. Acad. Sci. USA} \textbf{\bibinfo{volume}{101}},
  \bibinfo{pages}{3747} (\bibinfo{year}{2004}).

\bibitem[{\citenamefont{Serrano}(2007)}]{Serrano:2007}
\bibinfo{author}{\bibfnamefont{M.~A.} \bibnamefont{Serrano}},
  \bibinfo{journal}{J. Stat. Mech.} p. \bibinfo{pages}{L01002}
  (\bibinfo{year}{2007}).

\bibitem[{\citenamefont{Barth\'{e}lemy
  et~al.}(2003)\citenamefont{Barth\'{e}lemy, Gondran, and
  Guichard}}]{Guichard:2003}
\bibinfo{author}{\bibfnamefont{M.}~\bibnamefont{Barth\'{e}lemy}},
  \bibinfo{author}{\bibfnamefont{B.}~\bibnamefont{Gondran}}, \bibnamefont{and}
  \bibinfo{author}{\bibfnamefont{E.}~\bibnamefont{Guichard}},
  \bibinfo{journal}{Physica A} \textbf{\bibinfo{volume}{319}},
  \bibinfo{pages}{633} (\bibinfo{year}{2003}).

\bibitem[{\citenamefont{Almaas et~al.}(2004)\citenamefont{Almaas, Kov\'{a}cs,
  Vicsek, Oltvai, and Barab{\'a}si}}]{Almaas:2004}
\bibinfo{author}{\bibfnamefont{E.}~\bibnamefont{Almaas}},
  \bibinfo{author}{\bibfnamefont{B.}~\bibnamefont{Kov\'{a}cs}},
  \bibinfo{author}{\bibfnamefont{T.}~\bibnamefont{Vicsek}},
  \bibinfo{author}{\bibfnamefont{Z.~N.} \bibnamefont{Oltvai}},
  \bibnamefont{and} \bibinfo{author}{\bibfnamefont{A.-L.}
  \bibnamefont{Barab{\'a}si}}, \bibinfo{journal}{Nature}
  \textbf{\bibinfo{volume}{427}}, \bibinfo{pages}{839} (\bibinfo{year}{2004}).

\bibitem[{\citenamefont{Herfindahl}(1959)}]{HHIHerfindahl:1959}
\bibinfo{author}{\bibfnamefont{O.~C.} \bibnamefont{Herfindahl}},
  \emph{\bibinfo{title}{Copper Costs and Prices: 1870-1957}}
  (\bibinfo{publisher}{John Hopkins University Press},
  \bibinfo{address}{Baltimore, MD, USA}, \bibinfo{year}{1959}).

\bibitem[{\citenamefont{Hirschman}(1964)}]{HHIHirschman:1964}
\bibinfo{author}{\bibfnamefont{A.~O.} \bibnamefont{Hirschman}},
  \bibinfo{journal}{American Economic Review} \textbf{\bibinfo{volume}{54}},
  \bibinfo{pages}{761} (\bibinfo{year}{1964}).

\bibitem[{\citenamefont{Derrida and Flyvbjerg}(1987)}]{Derrida:1987}
\bibinfo{author}{\bibfnamefont{B.}~\bibnamefont{Derrida}} \bibnamefont{and}
  \bibinfo{author}{\bibfnamefont{H.}~\bibnamefont{Flyvbjerg}},
  \bibinfo{journal}{J. Phys. A} \textbf{\bibinfo{volume}{20}},
  \bibinfo{pages}{5273} (\bibinfo{year}{1987}).

\end{thebibliography}

\end{document}